# Internal Versus External Recruitment – The Story of Three Consecutive Project Managers in an IT Project

**Michelle Ye**
School of Engineering and ICT
University of Tasmania
Hobart, Australia
Email: yaqian.ye@utas.edu.au

**Kristy de Salas**
School of Engineering and ICT
University of Tasmania
Hobart, Australia
Email: kristy.desalas@utas.edu.au

**Nadia Ollington**
Faculty of Education
University of Tasmania
Hobart, Australia
Email: nadia.ollington@utas.edu.au

## Abstract

As project managers (PMs) play an important role in project success, assigning PMs with appropriate skills and personalities to projects is a crucial task. Nevertheless, empirical research on skill requirements for information technology (IT) PMs is limited and little information systems literature focuses on the role of internally recruited IT PMs. This paper presents a case study of a troubled IT project led by three consecutive PMs, with a range of backgrounds, skills, and personality types. Across subjects, IT project management was found to be a necessity of project success. Additionally, it was observed that internally recruited PMs showed advantages in understanding organisational culture and business processes. Lessons learned from the three PMs confirm the importance of particular skills previously described in the literature, and the need for an additional focus on how an IT PM's personality facilitates or inhibits IT project outcomes.

## 1   INTRODUCTION

In recent decades, Information Technology (IT) Project Management has played an important role in organisational studies (Braun et al. 2012; Gillard 2005; Kutsch 2008; Napier et al. 2009; Petter and Randolph 2009; Roy et al. 2010). IT project management, also known as software project management, is defined as "the process of planning, organising, staffing, monitoring, controlling and leading a software project" (IEEE Standards Board 1987, p. 10). With new systems and new technology to be embedded in organisations, IT projects are facing many challenges since they involve organisational changes to achieve effectiveness and efficiency in organisations (Avison and Torkzadeh 2008; Iacovou et al. 2009; Keil and Mahring 2010; Seddon et al. 2010; Standish Group 2004). The commonly identified factors of interest and influence regarding IT project success/failure include: project size and complexity; use of methodology; top management engagement or support; user liaison (or even user resistance); stakeholder communication; organisational politics; and other change management concerns (Gillard 2005; Grainger et al. 2009; Kappelman et al. 2006; Liebowitz 1999; Liu and Deng 2015; Liu et al. 2010; Oz and Sosik 2000; Reich et al. 2008). These factors are related to how to enable change by effectively managing the relationship between corporate management/users and the project, with little attention on IT Project Manager (PM) skills. Despite prolific advice being offered by research papers on the skills of successful PMs (Anantatmulla 2010; Gehring 2007; Pettersen 1991; Thamhain 1991; Thamhain and Gemmill 1974; Turner and Müller 2005), research on the role and skills of IT PMs and their influence on IT projects is limited (Napier et al. 2009).

There appears to be a focus within existing studies concerning IT PMs towards career PMs from IT consultancy companies (Jiang et al. 1998; Napier et al. 2009; Smith et al. 2011). In these studies, data collection is commonly based on career IT PMs' past experiences of multiple IT projects. In the current study, career IT PMs are PMs who have IT project management expertise (e.g. who have completed tertiary study in project management) or experience in managing large IT projects. Another group of interest are non-career IT PMs, also called 'accidental' PMs (Hunsberger 2011; Tarne 2003; Turner et al. 2003; Walker and Peterson 1999), who have not had professional project management experience, but after acquiring subject matter expertise and a good reputation within the organisation are promoted into the role of PM (e.g. internally recruited IT PMs). Internal non-career IT PMs tend to have influence and understanding of the corporate culture, and thus have potential advantages in leading IT projects. In information systems (IS) research, the role that non-career IT PMs play in affecting an IT project is lacking, especially the comparison between non-career and career IT PMs. In this light, the exploration of how non-career IT PMs lead IT projects, and the differences between career and non-career IT PMs, may add to our understanding of IT PMs skills and IT project management.

The objective of this paper is to shed light on the differences between internally recruited non-career IT PMs (who have a cultural advantage) and external career IT PMs (who have project management expertise and experience), and further, to provide insights into the competencies of IT PMs that affect IT projects positively. In order to achieve this objective, this paper reports on a case study of a troubled IT project led by three PMs consecutively, all with significantly different backgrounds, skill sets, and personalities. As a result, the three PMs encountered different challenges and resorted to different ways of managing the issues. By learning lessons from the stories of the three PMs' experiences, one can gain a deeper understanding of what is important to IT PMs for leading and managing a troubled IT project.

## 2　THEORETICAL FOUNDATIONS

With the rapid development of IT in today's organisational life, IT projects are widely embedded and IT project management is a popular topic in the discourse of organisational management (Avison and Torkzadeh 2008; Boonstra 2013; Heracleous and Barrett 2001; Morris 1997). Research has highlighted the critical role that IT PMs play in affecting an IT project's success or failure (Patanakul 2011; Petter and Randolph 2009; Smith et al. 2011).

This section presents a review of the literature concerning two major factors in relation to IT PMs that have impact on project progression. The two factors are the assignment of IT PMs and the competencies of IT PMs.

### 2.1　IT Project Manager Assignment

Assigning the right PM to an IT project is a crucial decision that corporate management needs to make. Research on PM assignment has been especially sparse and only a few studies draw a particular focus on this aspect and its impact on project management effectiveness (Patanakul 2011; Patanakul and Milosevic 2006; Patanakul et al. 2007; Pettersen 1991). Patanakul (2011) indicates that a good match between project requirements and the competency of a project manager is an important success factor of a project, which will further lead to business success. Therefore, it is critical to understand project requirements in terms of the capabilities required for leading a project. Previous studies have found that such requirements may include the natural characteristics of an IT project such as technological uncertainty (Nidumolu 1995), and system complexity (Shenhar 2001b), as well as environmental factors such as organisational culture (Elmes and Wilemon 1988), and resource allocation for multi-project management (Engwall and Jerbrant 2003). For example, Shenhar's (2001a) study found that in projects with low technological uncertainty, PMs usually act as administrators whose major goal is to complete the project as planned, while high technological uncertainty types of projects require highly qualified PMs and professionals for their technical and engineering skills (hard skills) in addition to their administrative skills (soft skills).

Shenhar (2001b) and Elmes et al. (1988) emphasise that hard and soft skills are important factors for determining the management style and management effectiveness of PMs. In this paper, PM hard skills are qualified technical competencies of IT project management concerning processes, procedures, tools, and techniques (Azim et al. 2010); whereas soft skills refer to leadership, managerial and interpersonal skills (e.g. understanding and gaining the trust and confidence of customers) (Agrawal and Thite 2006).

IT PMs with different backgrounds (e.g. externally recruited or internally assigned) may have advantages and disadvantages regarding their hard and soft skills of project management. In many cases, IT PMs are employed externally by a supplier organisation and have a duty to deliver on contract (Reich and Sauer 2010). An advantage of external IT PMs is their professional skill sets and experience of project management, especially the hard skills such as systems/business analysis skills (Green 1989), and project management processes, tools, techniques, and methodologies (Sukhoo et al. 2005). The limitations of employing external IT PMs may be that they usually have little understanding of the organisation's philosophy or of the culture of the organisation (Elmes and Wilemon 1988), and thus have little influence within the organisation. Internally recruited IT PMs, on the other hand, have the benefit of cultural and interpersonal influence and an understanding of corporate business processes. However, internally recruited IT PMs, since their role of PM is 'accidental', may lack essential project management skill sets and leadership competencies for the role (Gehring 2007; Walker and Peterson 1999). Given the distinct difference between career and non-career IT PMs, our study looks into the comparative role differences between internally and externally recruited IT PMs in order to provide insights into the knowledge of IT PM skills and effective project management. As the research suggests, the advantages of external career IT PMs (i.e. project management expertise and experience) and internally non-career IT PMs (i.e. cultural/interpersonal influence) are important IT PM requirements for project success (Napier et al. 2009; Sukhoo et al. 2005).

## 2.2　IT Project Manager Competencies

A number of publications have identified the skill sets of successful PMs (Anantatmulla 2010; Archibald 1992; El-Sabaa 2001; Frame 1999; Gaddis 1959; Pettersen 1991; Thamhain 1991); however, studies on IT PM competencies are limited, and only a few studies have a particular focus on the skills of IT PMs (Jiang et al. 1998; Napier et al. 2009; Petter and Randolph 2009; Smith et al. 2011). A relevant study reported by Jiang et al. (1998) aimed to identify a set of ranked IT PM skills, and found skills ranging from top level to least required. These included: interviewing; directing and managing; communication; interpersonal skills; and skills of sales assertiveness and non-verbal communication, respectively. A weakness in Jiang et al.'s study discussed by Napier et al. (2009) was that the questionnaire items were originally designed for examining the skills for system analysts, which might miss out on other possible important skills for IT PMs. As such, Napier et al. (2009) adopted an open-ended interview approach that allowed for the emergence of the most relevant IT PM skills, and found that aside from previously identified skills, there were other skills that could also impact IT project management effectiveness, such as client management, planning and control, and problem solving.

Another study concerning IT PM skills included a focus on soft skills to manage user expectations in IT projects (Petter and Randolph 2009). Petter and Randolph confirmed the importance of many PM soft skills previously described in Muzio et al.'s article (2007) (e.g. communication, organisational culture, leadership, problem solving, decision making, team building, flexibility and creativity, and trustworthiness). These authors also discovered the need for an additional focus on how social norms and organisational conditions encourage or inhibit sharing and re-using of knowledge regarding the soft skills needed to run effective IT projects.

In contrast to these studies, there are studies concerning IT PMs' soft competencies in relation to the managing of emotions, or their belief about personal competence (Jani 2011; Smith et al. 2011). For example, Smith et al.'s (2011) study presented stories of IT PMs' personal experience relating to optimism or stress, and how the soft competencies of managing optimism or stress influenced the project outcome. Their findings suggested that stress could positively and negatively influence IT project success, and that as such, IT PMs should have a positive but realistic degree of optimism to carefully manage project team stress. Their use of a storytelling method also provides an in-depth and rich understanding of how IT PMs can use their soft skills in the areas of optimism and stress management to improve project management. This is important because too much optimism or high stress levels are among the reasons why IT projects fail (Henderson 2006; Sethi et al. 2004). In another example, Jani (2011) implemented an experiment using a created scenario of a failing IT project and measures of self-efficacy and risk perception to investigate the relationship between PMs' self-efficacy, project risk factors, and commitment to a failing project. In this study, self-efficacy was defined as an individual judgement about how well they could perform in a particular task situation. It was found that, although IT PMs could accumulate sufficient experience to identify the necessary strategies for gaining control over a complex project, learning from past projects might not necessarily translate to novel projects, and further, past success could lead PMs to underestimate the risks of a troubled IT project. This was termed a 'self-efficacy bias'. Jani's finding is in opposition to a previous study by Du et al. (2007), which found that experienced PMs had higher risk perception compared to those with little expertise of project management. Importantly, these studies compared experienced PMs with students who either had gained some knowledge of the software development lifecycle through coursework (Jani 2011) or minimal knowledge of IT projects (Du et al. 2007) (i.e. novices), and therefore lacked a direct comparison between inexperienced PMs and experienced PMs.

A comparison of internally and externally recruited IT PMs is therefore important, since these different types of IT PMs have potential advantages and disadvantages regarding effective project management. It appears that to date, such a comparison is lacking in the literature. The current study presents the stories of three consecutive IT PMs with different backgrounds, skill sets and personalities, in the same IT project. The site for this research was a university in the Asia Pacific region named AsiaPac University (pseudonym), which involved the implementation of a large student system (SS) that integrates approximately 150 systems covering various business areas. This was a core business transformation project in AsiaPac University, and as such identified as being suitable for the current study.

## 3 RESEARCH METHOD

Based on the subjective ontology and interpretivist epistemology, a longitudinal (two-year) case study was conducted. The case study method was adopted because it is particularly well-suited for understanding the interrelationships between IT-related change and management practices in an organisational context (Benbasat et al. 1987; Darke et al. 1998; Doolin 1996; Kling and Iacono 1984).

### 3.1 Data Collection

The data collection techniques included semi-structured interviews, observation, and documentation study. Forty-eight semi-structured interviews (40-90 minutes in length) were conducted with 46 project stakeholders (see Table 1), as the primary source of data over a period of 2 years (from May 2012 to June 2014). Observation in project activities and examining project documents (e.g. Project Plans, Business Cases, Project Team Structure diagrams) were used to contextualise and confirm the researcher's understanding of data throughout the analysis phase. This was because using multiple sources of data is a key strength of the case study method (Yin 2009). As Yin (2009) noted, the adoption of multiple sources of data collection can minimise the researcher's bias and influences over the interpretation of the phenomena.

| Key Stakeholders | Responsibilities/Concerns |
|---|---|
| Senior Management (SM1-5) *(Deputy Vice Chancellors, Chief Operating Officer, Deans and Associate Deans of faculties)* | Increase the efficiencies of service delivery and the secure and controlled exchange of data. |
| | Finalise the implementation of the SS project with no more budget and timeline blowouts. |
| Business Administrative Group (BS1-6) *(Heads of Service in various business areas of the university)* | Comply with Senior Management's desire to try out the new product of systems. |
| | Provide critical business expertise and decision making within the scope of their role in support of the SS project. |
| IT Division Management Group (TM1-2) *(Chief Information Officer and the Associate Director in IT Division of the university)* | Provide IT services support to assist the implementation of the SS project. |
| | Increase credibility and recognition of IT Division from other areas of the university. |
| Transition Support Group (TO1-5) *(Transition support staff recruited for the SS implementation)* | Support major organisational change and help students and staff throughout the implementation and transition phases of the SS project. |
| Project Leaders (PL1-5) *(Project Managers/Directors, Assistant/Deputy Project Directors)* | Deliver the project against the budget and timeline with minimum risks and issues left to the post-implementation stage. |
| | Obtain more power and recognition of the importance of the SS project. |
| Project Middle Management (PM1-8) *(Project team stream leaders, team leaders, senior consultants, Communication Manager)* | Follow the project leader's leadership to manage deliverables/outcomes for their team/stream against the budget and timeline. |
| Business Analysts (BA1-8) *(Business Analysts in the project team)* | Follow their team/stream leader's instructions to capture business process, functional and technical requirements for the specification of project solutions. |
| IT Workers (TW1-5) *(System Programmers, System Developers, System Testers in the project team)* | Follow their team/stream leader's instructions to be responsible for a series of IT related tasks including requirements engineering, software design, development, testing and documentation. |
| Training Team (TT1-2) *(Training Developer, Training Agent)* | Provide training services to users all over the university for effectively running the implemented systems. |

*Table 1. Roles of the participants and their responsibilities/concerns*

*Note: The codes in brackets in the 'Key Stakeholders' column (e.g. SM1-5) refer to the labels assigned to each of the participants that were interviewed.*

Semi-structured interviews were chosen because there were specific topics that needed to be narrowed down and covered, but at the same time we wanted to hear the participants' stories (Rabionet 2011). In each interview, a number of general questions (approximately 2-3) were asked to start up the conversation in the interview. With some idea of the important issues that occurred in the IT project, the interviewer prepared additional questions to probe for information that may otherwise have been missed. Improvisation and listening strategies, as suggested by Myers and Newman (2007), were used in order to construct questions or provide prompts based on the participant's response. The interviews were delivered face-to-face, except for one telephone interview. Each of the 47 face-to-face interviews was recorded using a digital audio voice recorder, and written-notes were recorded by the interviewer during the telephone interview.

Due to the sensitivity of the current research focus, concerns were raised in relation to how the voices of the participants might be represented appropriately and ethically, and how to provide reliability and authenticity. In order to establish rigour and trustworthiness in the research process and therefore the findings, the current research drew upon the set of principles for evaluating interpretive field research proposed by Klein and Myers (1999), that is, taking into account participant feedback, the historical and social context, and participant's own interpretations, alongside researcher sensitivity via observation and documentation.

### 3.2  Data Analysis

Data was analysed using an adapted coding paradigm developed by Creswell (1998), based on grounded theory. Grounded theory principles assisted to provide the initial framework for the coding process (Glaser 1992; Glaser 1978), which included three conceptual coding levels: open, axial and selective. Open coding includes an inductive coding process from the raw data and axial coding uses the process of relating codes to each other until themes emerge. Once themes had emerged from the data categories and sub-categories, selective coding was carried out. This final phase included taking note of the social behaviours and activities that were in the coded data set, how the coded data represented the emerged themes, and further, how the themes and the data set could answer the research questions (Creswell 1998).

## 4  FINDINGS

The focus for this case study was upon the stories of three consecutive IT PMs, Bob, Michael and Sally (pseudonyms), during the implementation and institutionalisation of the SS project in AsiaPac University. The project was initiated in July 2006 and eventually went live in September 2014 (see Figure 1). During those eight years the project was troubled, since its first Business Case was approved in 2006 with multiple time and budget overruns. The Business Case was revised twice as was the go-live date. Until the end of our case study in 2014, the project had stalled in the planning stage (pre-implementation) of a project lifecycle (Project Management Institute 2004). Significant project structure changes also occurred, particularly to project leaders.

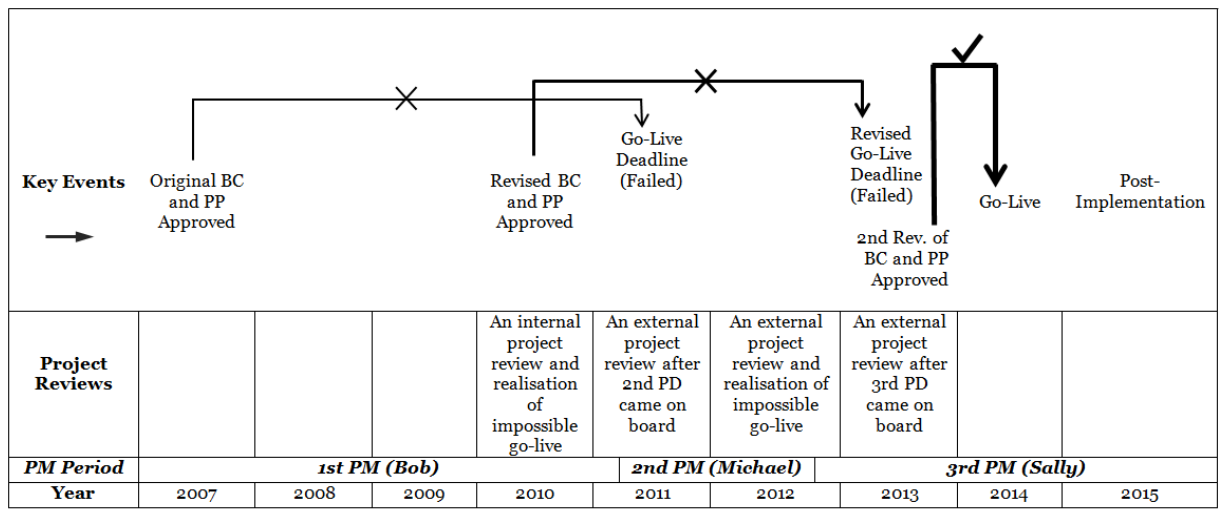

*Figure 1: The Timeline of the SS Project*

As shown in Figure 1, three PMs were involved in the project at different times. Bob, Michael and Sally came from different backgrounds, and had different skill sets and personalities. Accordingly, they encountered different challenges and problems, and resorted to different approaches for dealing with issues. All PMs were interviewed at least once. Bob and Sally were interviewed twice for further in-depth conversation as they were involved longer in the project compared to Michael. The stories around the three PMs are structured in the following three sections, one for each PM.

## 4.1　The University-Sourced Project Manager – Bob

Bob, the first PM of the SS project, led the SS project team for a period of four years. The project was allocated a budget of around $11 million and the go-live date for the SS system was June 2011. When the project team was initially assembled, it was a small team of 11 members seconded from a range of university business areas. The team was headed by Bob and assisted by an Assistant Project Manager and a Project Administrator. Other members of the team included four business/systems analysts as well as persons responsible for database administration issues, report generation, system testing, and end user training. As a previous senior administrator in the University prior to the secondment, Bob was considerably influential among other administrators. Moreover, he understood various aspects of the business process in the University. Therefore, even without IT project management experience, he was assigned by University Executives as the SS PM and worked closely alongside administrators in various business areas to formulate the key business requirements.

With more than 20 years of service to the University, a wealth of knowledge of organisational culture and business processes, and an 'extrovert' personality (as noted in interviews by his colleagues), Bob was a popular person. We refer to 'personality' in this paper as the distinctive character that one person shows to others, which can include a PM's 'role-play' in order to gain project success. The 'extrovert' personality that Bob displayed gave him a great advantage when he was attempting to engage business users during the requirements analysis stage. A business administrative manager and a transition support staff member working for the project expressed the cultural advantages that Bob had as follows:

*"I think [Bob] would definitely have understanding of the experience for the business in the University. He would definitely, in my opinion, he worked in [Student Services Centre], (and) he sees a lot happening in [Student Services Centre]. I think he would have quite good understanding of what happens around the university and how things are currently done. I am guessing that would inform how well communicative he was as a project leader that helped in communicating things. (…) I can probably see having a background in that culture could either be a positive or a negative, and certainly communication ways, understanding ways of how the university currently operates would absolutely be a benefit. (…) [Bob] has the skill to talk to the university. [Bob] has the skill to understand what the complexities are"* (a business administrative manager, BS4, lines 79-84; 89-91; 136-137, interviewed on 03 March 2014).

*"He (Bob) has the knowledge of the university, he understands how people work and think, and he could rally people around and smooth waters. I mean it's a very political environment as well, so he*

*was able to navigate that really well" (a transition support staff, TO1, lines 138-140, interviewed on 07 April 2014).*

Nonetheless, early progress was slow due to the fact that Bob did not have IT project management expertise or experience in managing large IT projects. As identified by externally recruited project specialists that came in later (and by Bob himself), the most salient issues plaguing the project under Bob's directorship included:

- poor business requirements elicitation and analysis – as commented by a senior business analyst who was a career business analyst recruited externally:

*"At one stage they had a thing called…, something like this, and it was represented from different areas so about 20 people from business so all sitting in circle, key people in meeting, and then looking at a particular area and determine what system that was going to be in, 'oh that's going to be in [a sub-system]'. So they didn't say 'what are our requirements', 'what do you try to achieve', 'how and what's the business process that can support our requirements and what's capability…' and I think it was a lack of experience in leadership mostly" (a manager of business analysts, PM4, lines 151-159, interviewed on 04 July 2013).*

- under-estimation of the budget, scope and timeline – as Bob said:

*"We were totally off the mark in terms of the initial business case in terms of the numbers required, the time required, and all the rest of it, totally off the mark, and that's based on the advice we received from a number of places so it wasn't what we just came up with, but we were totally off the mark and I take responsibility for that" (a project leader, PL1, second interview, lines 580-583, interviewed on 09 September 2013).*

- a lack of appropriate project skill sets in the project team – as the later PMs Michael and Sally said:

*"There was no change management expert or consultant on that team. They were talking about increasing the strength and increasing the skill base around change management and people doing change management things, but there was nobody on the team as far as I can make out who's done it in a large scale" (a project leader – Michael, PL3, lines 207-210; interviewed on 21 May 2012).*

*"I've got very few people with the same value as you need to have to run a project successfully because they are either they have never been on a project or they don't have the appropriate level of skills. So in other words, they've been asked to do 'this' job where they've actually got skills to do 'that' job, so that makes it extremely difficult" (a project leader – Sally, PL5, first interview, lines 521-524, interviewed on 14 December 2012).*

Another inhibitor, driven from directorship of the 'de facto' leaders who came from positions within the University was that the internally recruited project team was submissive to the organisational hierarchy, and thus they tended to convey 'better-than-real' news to the University Executives. A project consultant later commented on this issue:

*"I think if they come from this university, they're immediately subordinate to the traditional corporate hierarchy here rather than the project hierarchy, so they don't know how to counter or challenge, understandably, for all organisational power. If you come from other universities, as a contractor, you will be more comfortable about that. If you come from a vendor or service provider, a business like ours, then you assume you have to confront and in some ways challenge the negotiated order or the world will be as it was" (a change consultant, PM1, lines 454-459; interviewed on 28 February 2013).*

As a result, even though early progress was slow and some problems emerged, it was assumed that the project was progressing reasonably well, until a highly critical review of the project caused considerable consternation to the University Council in 2010. Discussion among Council members and the Vice Chancellor led to the Vice Chancellor replacing Bob after 4 years leading the project team. Bob was removed from the project and returned to his administrative role in the University, acting as a senior subject matter expert instead of a project leader. For Bob, this was a positive move forward for the project, and he suggested that an external IT PM would be a better fit for delivering the remaining aspects of the project. Nevertheless, it seemed to many in the University, including those in the SS project team, that the removal of Bob was perhaps not accompanied by a rigorous level of analysis. It seemed to some in the project team that senior management had 'stamped its foot' in anger, but had not diagnosed the problems that were plaguing the project.

## 4.2　The 'CEO'-Style Project Manager – Michael

The second PM of the SS project, Michael was recruited from outside the University. Michael had been the CEO of an online betting company, and hence he was assumed to have a reasonable business-oriented knowledge of IT systems. Critically, however, he had no experience in directing and managing large and complex IT projects. Michael reviewed the original Business Case and had a new Business Case prepared. This second Business Case was approved in December 2010 with a new timeline, and an increased budget and resource. The new project budget almost doubled to $22.7 million and the project team was increased to over 30 members. Michael also made significant changes to the project team structure, and the go-live date was changed to April 2013. Nevertheless, the project fared little better under Michael.

Michael had a remote 'CEO style' and did not tend to set clear directions for the project team leaders. His lack of hands-on project management became obvious to a number of senior members of the project team. Several issues were identified by project team members and business stakeholders regarding Michael's management style:

- a lack of hands-on, a remote management style – as expressed by a project worker:

*"[Michael] was disinterested, remote, didn't really matter. If you made a good point about something he could just override it anyway. You know, it's just… Remote is a pretty good word" (a Business Analyst, BA7, lines 122-124; interviewed on 09 May 2014).*

- a lack of both cultural understanding and project management expertise. As an externally recruited PM, Michael was perceived by the university staff as an 'outsider', who they did not identify with. This made it worse for Michael to improve his project management – a business administrative manager expressed this issue:

*"[Michael] certainly didn't have an understanding of higher education, so I think that, that was also a challenge because you need to have that background, (and) that experience. I mean, [Bob] at least had the culture, you know what I'm saying, the cultural understanding and the organisational structure and all those things. To not have both is… hmmm" (a business administrative manager, BS1, lines 310-314, interviewed on 01 March 2013).*

- poor communication. In particular, Michael's communication towards the senior management team in the University was reassuring, and so it caused further consternation to the Council and Vice Chancellor when a second external review of the project showed it to be in a parlous state, having no chance of meeting the planned go-live date. The second PM Michael, seeing the writing on the wall, left the University.

## 4.3　The Project Manager with Strong Personality – Sally

In late October 2012, the third PM, Sally, who had experienced complex IT projects was recruited from outside the University. Sally as a new PM, needing to accurately and publicly establish the current state of the project, instituted another comprehensive external review of the project and then based on the findings, argued that the April 2013 go-live target was no longer possible. However, the senior management of AsiaPac University were now becoming impatient with the retreating go-live dates, and were also becoming increasingly aware of, and frustrated with, the inadequacies of the current legacy system and the attendant ineffective and inefficient business processes. Thus the Vice Chancellor brought considerable pressure to bear on Sally, insisting among other things, that she meet the deadline of September 2013 as a compromise of the previous April deadline.

This time however, Sally was a highly experienced PM who could clearly see that the deadline was completely unrealistic and so refused to commit to it. This caused a 'battle' between Sally and University Senior Executives. For a while, Sally confronted the University Senior Executives by almost walking out of the position. During this time of uncertainty for the SS project, the Vice Chancellor requested that the University CIO get involved in the project to assess whether the deadline was in fact unattainable as Sally was asserting. The CIO spoke with the project team members and was told quite firmly by a number of them that Sally was correct. Sally then returned to her role as head of the project and presented the whole picture of the project to the University Council, which caused another scare. The University Council chose to trust Sally's expertise and agreed with Sally's assessment of July 2014 as a go-live date for the implementation of the new system.

Sally then started detailed work on the new Project Management Plan and the third Business Case based on the revised timeline, resource, and budget, at the same time working with the external review team and re-structuring the project team. By that time, the new project budget almost doubled again to $40 million, and the size of the project team was significantly increased to over 60 members. The SS project then began to move in the right direction under Sally's leadership and finally went live within the designated timeline.

Overall, there were several reasons that emerged and were identified by various project stakeholders that could explain why Sally succeeded in persuading the University Senior Executives and making the project live. The following characteristics determined Sally's success in the SS project:

- project management expertise and experience – as a university senior manager said:

*"There is a very strong respect from the two [Project Sponsors] of [Sally's] capability, and there is very clear understanding from them that she holds a key to pulling this off. And there has been at least one other occasion where [Sally] very basically had to advise to two [Project Sponsors] by saying that 'you've got to look after this, you've got to deal with this matter, this is my recommendation how you deal with it', because if you don't, you are under the risk of [Sally] walking. If [Sally] walks, this goes down the gurgler, and that has been a clear understand. They know that, they know that all to full well, so it is a very healthy respect that they have of [Sally's] capability as a Project Director" (a senior management member, SM5, lines 178-186; interviewed on 30 October 2013).*

- communication and persuasion skills – as an early senior project staff said:

*"She (Sally) is very very good at building arguments, persuading people, just from the little I've seen. I wouldn't be surprised if we don't run out of money and she gets there and she says 'I have an argument as to why we should do this and it's going to cost this much more but because there is only that much to go, so I can give you this much more certainty we would go'. But you have to have that, you have to have somebody who's got that mind and that ability to put that forward, to actually do it. It's the message we've got for years: 'don't you dare come back for another cent'" (a project leader, PL2, lines 450-456; interviewed on 09 May 2013).*

- an 'assertive' personality with seemingly high self-efficacy – as Sally described she had to be strong and firm with her decision and approach:

*"I've had to come in and be very firm, very belligerent on my most of the point of 'no I'm not accepting that'. So it's being strong rather than…" (a project leader, PL5, first interview, lines 262-263; interviewed on 14 December 2012).*

Further, Sally continued to express that Bob's personality was overly consultative, and possibly among the inhibitors of the project:

*"What I hear is they've done four reviews, and taken action on none of the actions. A classic one was [Bob] when he was with this thing. He is a classic, he is a NF dreamer, Myers-Briggs personality profile type. So he wants to please people and he does. He will always be like that. He would consult, consult and consult. But I want decisions like this. I am already at the answer 5 minutes before he's even thought of half of the questions, but I have to go the journey with him, so it's me understanding that's his style. But as a Director, he would not have known what was hitting him in and I think they shot themselves in the foot. They had the wrong idea of having a business person managing the programme" (a project leader, PL5, second interview, lines 473-481; interviewed on 16 September 2013).*

In summary, this section has presented the stories from the three consecutive IT PMs' experiences in the case study. The following section discusses lessons that we can learn from the three IT PMs' stories.

## 5　LESSONS LEARNED

Our findings from these stories confirm the importance of particular IT PM skills that have been previously described in the literature (Grainger et al. 2009; Kappelman et al. 2006; Liebowitz 1999; Oz and Sosik 2000), including hard skills, in particular project management expertise, skills for business requirements elicitation and analysis, project planning skills, and soft skills, in particular hands-on management style, communication, and persuasion skills. Moreover, we discovered the need for the additional focus on how an IT PM's personality encourages or inhibits the IT project outcome. This is in line with Gehring (2007) and Turner et al.'s (2003) findings that personality traits are important in influencing project management generally. Specifically, we found that being overly consultative could lead to a lack of progress for the project as a whole. For example, an IT PM with a wealth of project management expertise or experience tends to be confident and firm in his or her decision/judgements, and thus tends to give clear direction to the project team, leading to project success. Though previous research has found that IT PMs with high self-efficacy tended to underestimate project risks (Jani 2011), in our study the IT PM with seemingly high self-efficacy gained project success, perhaps because it was supported by strong project management expertise and experience, and an assertive personality. Therefore, having strong project management expertise and experience alone may perhaps not be sufficient. It may be important to combine these with high self-efficacy. In the current project, these characteristics contributed to an essential skill set of the IT PM, especially in a troubled, complex IT project with a tight budget and timeline.

One can infer from the first PM (Bob's) experience that an internally recruited (accidental) PM had advantages in leading the project because of his cultural understanding and business process knowledge. However, without any project management expertise or experience, the PM could not lead the project in an effective way because he and his 'de facto' (i.e. internally recruited) staff group largely underestimated the size and complexity of the project, and as such the resources required (i.e. costs, time, and skill sets). This 'de facto' project leadership did not have the necessary expertise to support staff to effectively elicit and analyse the business requirements or to identify the types of skill sets needed in project recruitment. Hence, in this case, the potential advantages of an internal PM seemed insufficient for the delivery of a complex IT project with project management expertise a necessity for project success. This finding supports Gehring's (2007) suggestion that PM assignment should be based on both leadership ability and technical expertise.

Therefore, an organisation's management should not rely on an 'accidental' PM without the necessary project management skill set to lead a complex IT project. Rather, as internal PMs have cultural influence, which we have shown potentially aids organisational change by encouraging business engagement, we suggest that a more suitable leadership role would be in leading the change that the project will bring to the organisation, for example a project champion or change champion (Morton 1983). Indeed in the case study, the SS project relied on the accidental PM (Bob's) cultural knowledge and influence within the University to foster the organisational change to be acknowledged and accepted during the whole 6-year pre-implementation stage, even after he returned to his administrative role from the project.

There are limitations within this study. First we focused on three individuals, which offered a limited scope. The second is in relation to the strength of our findings. For example, our assumptions based around self-efficacy are somewhat weak, particularly considering self-efficacy was not measured. Nevertheless, these findings highlight the importance of further work in this area. For example, future work should aim to incorporate a survey approach to measure personality traits including self-efficacy, in order to tease out possible relationships between personality and IT PM performance. A third limitation is that with its focus towards internal versus external recruitment, the current paper ignored a large number of 'other' factors inhibiting the SS project that had been identified, such as lack of senior engagement and support (i.e. poor project sponsorship), user reluctance and resistance to change, and organisational cultural impacts. These critical issues could form the basis for future research concerning project success and failure. Indeed, it would seem important to also explore these factors alongside personality factors. A fourth limitation of the paper is that the case study was conducted mainly during the pre-implementation stage of the SS project and ended at the project go-live stage. This made it difficult to compare required PM skills at different project stages. Future work could look into this aspect as different project phases often have different requirements on PM skill sets. Moreover, future research into studying different types of IT projects (e.g. non-university, with low technological uncertainty or less system complexity) could be helpful for evaluating our findings.

# 6   CONCLUSION

This paper aimed to provide insights into IT Project Manager (PM) skills, particularly focusing on the different roles that internally and externally recruited IT PMs (i.e. accidental versus career IT PMs) play in an IT project. We have learned lessons from three consecutive IT PMs' experiences within a complex, troubled IT project. Aside from commonly agreed essential IT PM skills such as project management expertise, good management style and communication tactics, our study has found that personality characteristics need additional focus in IT project management studies. We also found that the potential cultural advantage of internally recruited IT PMs could be utilised to enable and lead the organisation change triggered by the system implementation. In this light, we suggest internal IT PMs can be good project champions.

The paper has provided contributions to the Information Systems discipline at the theoretical and practical level. At the theoretical level, we have added to the knowledge of IT PM skills by presenting, and suggesting the need for additional focus on how personality traits and assignment of internally recruited IT PMs may affect IT project outcomes. At the practical level, the suggestions and lessons learned from the different types of IT PMs' stories from the case study have added to the knowledge concerning effective project management in IT projects. We encourage future studies to be conducted to investigate the relationship between different personality traits/types and IT PM success, and difference of skill and personality requirements for being a project champion and being an IT PM.

# 7　REFERENCES